\documentclass[aps,pra,superscriptaddress,amsmath,amssymb,preprintnumbers,twocolumn,showpacs]{revtex4}
\usepackage{amssymb}
\usepackage{graphicx}
\usepackage{bm}
\usepackage{color}

\newcommand{\Ignore}[1]{}

\newcommand{\Ket}[1]{\left\vert #1\right\rangle}
\newcommand{\Bra}[1]{\left\langle #1\right\vert}

\newcommand{\KetBra}[2]{\left\vert#1\right\rangle\left\langle#2\right\vert}

\renewcommand{\eqref}[1]{(\ref{#1})} 

\def\LZ{LZSM }

\begin{document}
\title{Microscopic description of dissipative dynamics of a level crossing transition}

\author{M. Scala}
\affiliation{Dipartimento di Fisica dell'Universit\`{a} di Palermo, Via Archirafi 36, 90123 Palermo, Italy}

\author{B. Militello}
\affiliation{Dipartimento di Fisica dell'Universit\`{a} di Palermo, Via Archirafi 36, 90123 Palermo, Italy}

\author{A. Messina}
\affiliation{Dipartimento di Fisica dell'Universit\`{a} di Palermo, Via Archirafi 36, 90123 Palermo, Italy}

\author{N. V. Vitanov}
\affiliation{Department of Physics, Sofia University, 5 James Bourchier blvd, 1164 Sofia, Bulgaria}

\begin{abstract}
We analyze the effect of a dissipative bosonic environment on the
Landau-Zener-St\"uckelberg-Majorana (LZSM) level crossing model by
using a microscopic approach to derive the relevant master
equation. For an environment at zero temperature and weak
dissipation our microscopic approach confirms the independence of
the survival probability on the decay rate that has been predicted
earlier by the simple phenomenological LZSM model. For strong
decay the microscopic approach predicts a notable increase of the
survival probability, which signals dynamical decoupling of the
initial state. Unlike the phenomenological model our approach
makes it possible to study the dependence of the system dynamics
on the temperature of the environment. In the limit of very high
temperature we find that the dynamics is characterized by a very
strong dynamical decoupling of the initial state ---
temperature-induced quantum Zeno effect.
\end{abstract}


\maketitle


\section{Introduction}

The exactly soluble Landau-Zener-St\"uckelberg-Majorana (LZSM) model \cite{ref:LZSM-1,ref:LZSM-2,ref:LZSM-3,ref:LZSM-4} is a very popular tool for estimating the transition probability between two quantum states whose energies cross in time.
A level crossing in combination with adiabatic evolution is of great practical significance for it leads to complete population transfer between the two states;
 various experimental level crossing techniques have been developed and demonstrated over the years \cite{ref:ARPC}.
Insofar as a quantum system is often immersed in a noisy environment,
 a number of studies have dealt with the effects of dissipative environments on this model \cite{ref:Ao89, ref:Akulin, ref:Vitanov97, ref:Nelbach, ref:Orth}.

Despite its extreme simplicity --- linearly changing energies and a constant coupling --- the \LZ model often provides unexpectedly
 accurate results when applied to real physical systems with elaborate time dependences.
This feature, and the appealing simplicity of the transition probability, have made this model vastly popular.
There are several additional intriguing features associated with the \LZ model.
It turns out that some standard \textit{approximate} methods, e.g. the Dykhne-Davis-Pechukas approximation \cite{ref:DDP}, when applied to this model give the \textit{exact} transition probability.
Moreover, when the upper state is allowed to decay irreversibly to other levels outside the two-state system with a decay rate $\Gamma$ the survival probability,
 i.e. the population of the lower state, does \textit{not} depend on $\Gamma$ \cite{ref:Akulin,ref:Vitanov97}, while the population of the upper decaying state vanishes at large times.
This is surprising because the evolution of the ground-state population does depend on $\Gamma$: it approaches its asymptotic value smoothly for nonzero $\Gamma$ and in an oscillatory manner for $\Gamma=0$.
It is also remarkable that an adiabatic elimination of the decaying state provides the exact result for the ground-state population.
It has been demonstrated that all these features are unique for the \LZ model and do not apply to any other model \cite{ref:Vitanov97}.

The latter results for the lossy \LZ model have been obtained in the framework of the phenomenological inclusion of the decay rate
 in the Schr\"odinger equation as an imaginary energy term in the Hamiltonian \cite{ref:Akulin,ref:Vitanov97}.
This approach is known to be justified for weak decay.
We have demonstrated recently that a rigorous microscopic master equation treatment, developed in Ref.~\cite{ref:Davies1978},
 can deliver dramatically different results compared to the phenomenological approach in the description of the stimulated Raman adiabatic passage technique
 in open three-state quantum systems in the regime of strong decay \cite{ref:Scala2010, ref:Scala2011}.
It is therefore interesting to examine the behavior of the \LZ model in an open quantum system by the master equation approach and verify its properties,
 e.g., the validity of the independence of the survival probability on the decay rate.
Moreover, the microscopic approach allows us to study the \LZ model in the regime of nonzero bath temperature, which is inaccessible by the phenomenological approach.
In view of the numerous applications of the \LZ model and the fact that a quantum system is often subjected to decoherence,
 our results are of potential significance in a number of physical systems \cite{ref:Nori_review},
 such as Josephson junctions \cite{ref:JJ1,ref:JJ2}, cold atoms in optical lattices \cite{ref:BEC}, spinorial Bose-Einstein condensates \cite{ref:Nori_spinorial}, and others.

The paper is structured as follows.
In the next section we present the physical system, by starting from the ideal case and then introducing the interaction with a bosonic environment and deriving the relevant master equation.
In the third section we compare numerical simulations obtained by our microscopic approach with predictions coming from the previously studied phenomenological model.
Furthermore, the dependence of the dynamics on the temperature of the environment is simulated and explained by invoking the notion of dynamical decoupling.
Finally, in the last section, we give some conclusive remarks.

\section{Physical System}

\subsection{The Ideal Case}

The ideal model comprises two quantum states $\Ket{1}$ and
$\Ket{2}$ coupled by the time-dependent Hamiltonian
\begin{equation}
  H(t) = \frac{\Delta(t)}{2}\,\sigma_z + \Omega(t)\,\sigma_x\,.
\end{equation}
In order to model the interaction of this system with a
dissipative environment, we add a third state $\Ket{3}$. Therefore
we consider a three-state system described by the Hamiltonian
($\hbar=1$)
\begin{equation}\label{eq:Hs}
H_{\text{S}}(t)= \left[\begin{array}{ccc}
-\Delta/2 & \Omega & 0 \\
\Omega & \Delta/2  & 0 \\
0 & 0 & -\omega_3
\end{array}\right]\,.
\end{equation}
We assume this model to be \lq\lq exact\rq\rq: this is possible
for instance if the two states $\Ket{1}$ and $\Ket{2}$ correspond
to two Zeeman sublevels of an atom or a molecule splitted by a
time-dependent magnetic field, while the coupling $\Omega$ is
induced by a static electric field in the presence of a dipole
operator which has only real matrix elements, i.e., $\vec{d}
(\Ket{1}\Bra{2}+\Ket{2}\Bra{1})$. It is worth noting that, in the
derivation of the Hamiltonian for such a system, no rotating wave
approximation is performed.

The eigenvalues of $H_{\text{S}}$ are $\pm \epsilon$, with
$\epsilon = \sqrt{\Omega^2+\Delta^2/4}$, and $\omega_3$. The
corresponding eigenstates are
\begin{subequations}\label{eq:eigenstates}
\begin{align}
\Ket{+}&= \cos\varphi\Ket{1}+\sin\varphi\Ket{2}\,,\\
\Ket{-}&= -\sin\varphi\Ket{1}+\cos\varphi\Ket{2}\,,\\
\Ket{3}& \,,
\end{align}
\end{subequations}
where $\tan\varphi = (\Delta/2 + \epsilon)/\Omega$. The quantities
$\Delta$ and $\Omega$ are, in general, time-dependent. In the
following we will focus on the case $\dot\Omega = 0$ (i.e., static
electric field) and $\Delta = \kappa^2 t$ (i.e., linearly changing
magnetic field), with $t$ spanning the range $[-\tau,\, \tau]$.

According to the calculations made in
\cite{ref:LZSM-1,ref:LZSM-2,ref:LZSM-3,ref:LZSM-4}, where the time
of interaction $\tau$ is assumed to be infinite, when the system
is prepared in state $\Ket{1}$  at $t=-\infty$, the survival
probability of state $\Ket{1}$ at $t=\infty$ is given by
\begin{equation}\label{eq:LZ formula}
  P_1(\infty)=e^{-2\pi\Omega^2/\kappa^2}\,.
\end{equation}
In the adiabatic limit, $\Omega/\kappa\gg 1$, the population of
the initial state $\Ket{1}$ vanishes, $P_1(\infty)\to 0$, and
complete population transfer $\Ket{1}\to \Ket{2}$ takes place.
This formula is approximately valid also for finite values of
$\tau$ provided $\tau\gg\Omega/\kappa^2$ \cite{ref:Vitanov96},
because the coupling $\Omega$ induces transitions between
$\Ket{1}$ and $\Ket{2}$ only in the proximity of the level
crossing. In other words, the \LZ formula \eqref{eq:LZ formula}
applies to finite times $\tau$ which are much larger than the
characteristic transition time, which in the nearly-adiabatic
regime is of the order of $\Omega/\kappa^2$ \cite{ref:Vitanov99}.
The values of $\tau$ considered in the numerical simulations below
will always satisfy the above condition.

\subsection{The microscopic dissipative model}

Let us now consider the interaction with a bosonic bath, so that the system is described by
\begin{subequations}
\begin{equation}
H = H_{\text{S}}+H_{\text{B}}+H_{\text{I}}\,,
\end{equation}
where
\begin{align}
H_{\text{B}} &= \sum_k \omega_k a^\dag_k a_k\,,\\
\label{eq:HInt}
H_{\text{I}} &= \left(\Ket{2}\Bra{3}+\Ket{3}\Bra{2}\right)\otimes\sum_{k}g_k\left(a_k+a_k^\dag\right)\,.
\end{align}
\end{subequations}
The interaction Hamiltonian in \eqref{eq:HInt} has the form
\begin{subequations}
\begin{equation}\label{eq:HInt_General}
H_{\text{I}} = A \otimes B\,,
\end{equation}
with
\begin{align}
A &= \Ket{2}\Bra{3}+\Ket{3}\Bra{2}\,,\\
B &= \sum_{k}g_k\left(a_k+a_k^\dag\right)\,.
\end{align}
\end{subequations}
Following the general theory of Davies \cite{ref:Davies1978} and the relevant consolidated approach \cite{ref:Florio2006, ref:Scala2010, ref:Scala2011}, the master equation can be written down, by using standard methods \cite{ref:Gardiner, ref:Petru}, in terms of the instantaneous jump operators between the eigenstates \eqref{eq:eigenstates} of the system Hamiltonian,
\begin{subequations}
\begin{align}
\label{eq:APiuMeno} A(\omega_{+-}) &= A_3(\omega_{-+}) = 0\,, \\
\label{eq:APiuTre} A(\omega_{+3}) &= \sin\varphi \KetBra{3}{+}\,,\\
 A(\omega_{3+}) &= \sin\varphi \KetBra{+}{3}\,,\\
\label{eq:AMenoTre} A(\omega_{-3}) &= \cos\varphi \KetBra{3}{-}\,,\\
 A(\omega_{3-}) &= \cos\varphi \KetBra{-}{3}\,.
\end{align}
\end{subequations}
Moreover, since we are in the presence of a level crossing we may
not be allowed to perform the rotating-wave approximation.
Therefore the master equation in the Schr\"odinger picture is
\begin{align}\label{eq:MasterEq}
\nonumber \frac{\mathrm{d}\rho}{\mathrm{d}t} &= -i\left[H_{\text{S}}, \rho \right] + \sum_{\omega,\,\omega'}\,\,\, \Gamma(\omega) [A(\omega)\rho\, A^\dag(\omega') \\
&- A^\dag(\omega') A(\omega) \rho ] + \mathrm{H.c.}\,,
\end{align}
with
\begin{equation}
\Gamma(\omega) = \int_0^\infty \mathrm{d}s\,e^{i\omega s} \mathrm{tr}_B\left[B^\dag(t) B(t-s) \rho_B(0)\right]\,,
\end{equation}
where $B(t)$ is the operator $B$ in the interaction picture at time $t$.
In the continuum limit we have
\begin{equation}
\Gamma(\omega) = \left|g(|\omega|)\right|^2 D(|\omega|) (1+N(\omega))\,\frac{\omega}{|\omega|}\,,
\end{equation}
where $g(\omega)$ is the coupling constant $g_k$ corresponding to the mode of frequency $\omega$, $D(\omega)$ is the density of the bath modes at frequency $\omega$, and $N(\omega)$ is the number of photons in each mode at frequency $\omega$.
Since the latter quantity is the occupation number for a thermalized boson system: $N(\omega) = 1/(e^{\omega/\Theta}-1)$, where $\Theta=k_B T$ with $T$ being the temperature
and $k_B$ the Boltzmann constant, the sign of $\omega$ is automatically taken into account since when $\omega<0$ one has $1+N(\omega)=-N(|\omega|)$.

\section{Asymptotic population of the ground state} \hskip0.5cm

\subsection{General behavior}

We have integrated numerically the master equation \eqref{eq:MasterEq} in order to find the behavior of the survival probability of the state $\Ket{1}$ after the interaction.
We consider for simplicity a flat spectrum for the reservoir, i.e. we assume $|g(|\omega|)|^2 D(|\omega|)\equiv \Gamma$ independently of $\omega$.

\begin{figure}[tb]
\includegraphics[width=0.40\textwidth, angle=0]{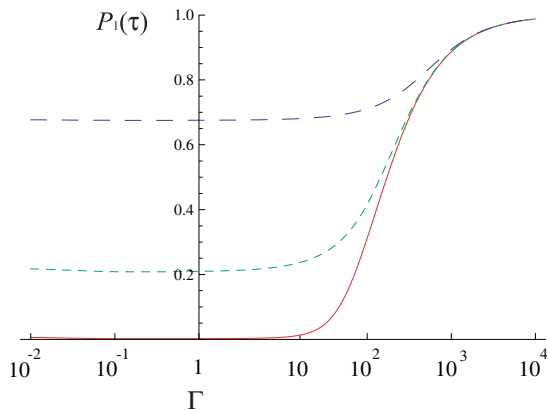} %
\caption{(Color online). Population of the state $\Ket{1}$ at
$t=\tau$ vs the decay rate $\Gamma$ (in units $\Omega$) at
$\Theta=0$, for different values of the chirp parameter $\kappa$:
$\kappa/\Omega=1$ (red solid line), $\kappa/\Omega=2$ (green
dashed line), $\kappa/\Omega=4$ (blue long dashed line). Here we
have assumed $\omega_3/\Omega=10^3$. Each time evolution occurs in
the time interval $[-\tau, \tau]$, with $\tau=30\Omega^{-1}$. The
initial state of the system is $\rho(-\tau)=\Ket{1}\Bra{1}$.}
\label{fig:Pop_Gamma_T0}
\end{figure}

In Ref.~\cite{ref:Vitanov97} a phenomenological model describing
the level crossing transition in the presence of losses toward
external states has been studied by adding an imaginary part
($-i\Gamma$) to the energy of the state $\Ket{2}$. The analysis
there and previous work \cite{ref:Akulin} showed that the survival
probability in the \LZ model does not depend on the decay rate at
all. However, in the \emph{finite} version of the \LZ model where
the interaction duration is finite \cite{ref:Vitanov96}, and in
several other analytic models, the phenomenological approach
predicts an increase of the survival probability in the presence
of strong damping (dynamical decoupling): $P_1(\infty)\rightarrow
1$ when $\Gamma\rightarrow \infty$.

Our approach gives essentially the same zero-temperature dynamics
as the one predicted by the phenomenological model for the finite
LZSM transitions. Figure \ref{fig:Pop_Gamma_T0} shows $P_1(\tau)$
(for $\tau\gg \Omega/\kappa^2$) as a function of the decay rate
$\Gamma$, for different values of the chirp parameter $\kappa$. As
$\kappa$ increases, the adiabaticity condition $\Omega/\kappa\gg
1$ deteriorates and more population is left in the initial state
at $\Gamma=0$. The figure clearly demonstrates that for low decay
rates the survival probability is independent of $\Gamma$, while
it increases for larger values of $\Gamma$. In the strong damping
limit the survival probability approaches unity as in the
phenomenological model.

\begin{figure}[tb]
\includegraphics[width=0.40\textwidth, angle=0]{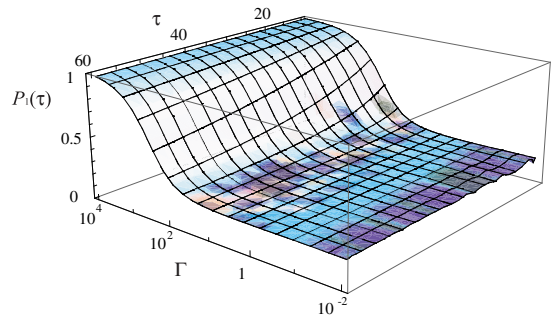} %
\caption{(Color online). Population of the state $\Ket{1}$ at
$t=\tau$ as a function of the duration $\tau$ (in units of
$\Omega^{-1}$, spanning the range $[10\,\Omega^{-1},
60\,\Omega^{-1}]$) and of the decay rate $\Gamma$ (in units of
$\Omega$). The dependence of the final population on the decay
rate is not affected much by the duration.}
\label{fig:Pop_Gamma_T0_Tau}
\end{figure}

Figure \ref{fig:Pop_Gamma_T0_Tau} shows the dependence of
$P_1(\tau)$ on both $\Gamma$ and $\tau$, revealing that the
dependence of the final population on the decay rate is only
slightly affected by the duration time $\tau$, and that for
smaller values of $\Gamma$ the influence of $\tau$ is even less
visible. However, we note that, increasing the value of $\tau$,
there is a slight shift of the region where the final population
starts increasing with respect to $\Gamma$. Since the larger
$\tau$ the larger  the value of $\Gamma$ needed to see a
significant increasing of the population $P_1(\tau)$, one can
conjecture that in the limit $\tau\rightarrow\infty$ one can
recover the independence of the population from $\tau$ predicted
by the phenomenological model. Nevertheless, developing this
analysis is beyond our present numerical resources.

\begin{figure}[tb]
\includegraphics[width=0.40\textwidth, angle=0]{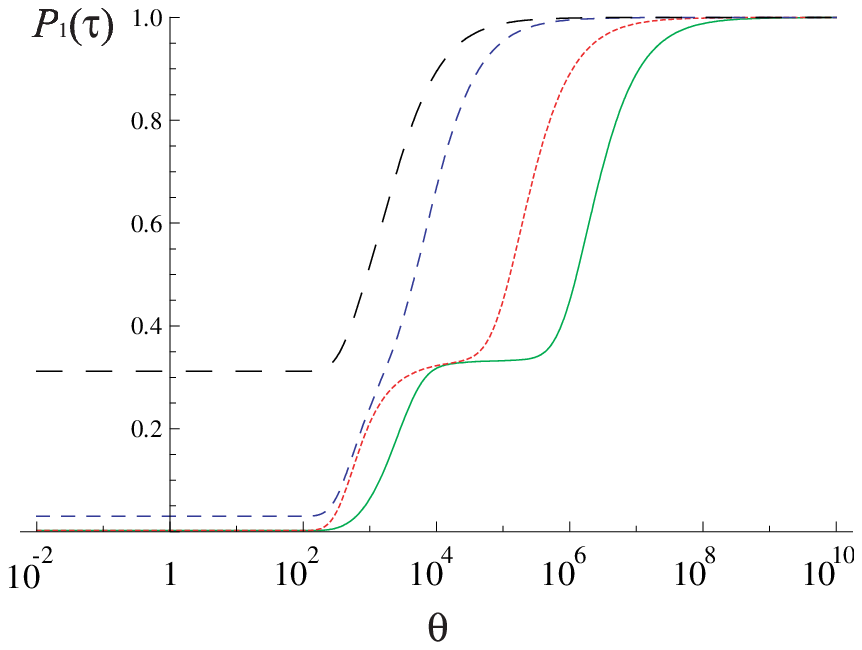} %
\caption{(Color online). Population of the state $\Ket{1}$ at
$t=\tau$ vs $\Theta$ (in units $\Omega^{-1}$), for
$\Gamma/\Omega=0.1$ (green solid line), $\Gamma/\Omega=1$ (red
dotted line), $\Gamma/\Omega=25$ (blue dashed line),
$\Gamma/\Omega=100$ (black long dashed line). Here we have
$\kappa/\Omega=1$, $\omega_3/\Omega=10^3$. Each time evolution
occurs in the time interval $[-\tau, \tau]$, with
$\tau=30\Omega^{-1}$. The initial state of the system is
$\rho(-\tau)=\Ket{1}\Bra{1}$.} \label{fig:Pop_Gamma_Tv1}
\end{figure}

\begin{figure}[tbh]
\includegraphics[width=0.40\textwidth, angle=0]{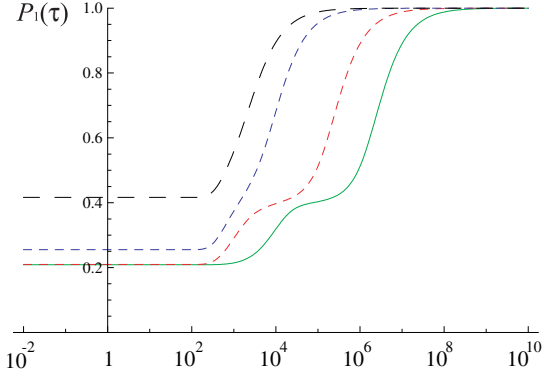} %
\caption{(Color online). Population of the state $\Ket{1}$ at
$t=\tau$ vs $\Theta$ (in units $\Omega^{-1}$), for
$\Gamma/\Omega=0.1$ (green solid line), $\Gamma/\Omega=1$ (red
dotted line), $\Gamma/\Omega=25$ (blue dashed line),
$\Gamma/\Omega=100$ (black long dashed line). Here we have
$\kappa/\Omega=2$, $\omega_3/\Omega=10^3$. Each time evolution
occurs in the time interval $[-\tau, \tau]$, with
$\tau=30\Omega^{-1}$. The initial state of the system is
$\rho(-\tau)=\Ket{1}\Bra{1}$.} \label{fig:Pop_Gamma_Tv4}
\end{figure}

Differently from the phenomenological approach which works at zero
temperature, our approach is able to predict the behavior of the
system also at nonzero temperature. Figures
\ref{fig:Pop_Gamma_Tv1} and \ref{fig:Pop_Gamma_Tv4} show the
survival probability as a function of the temperature for
different values of $\Gamma$. The two figures refer to different
values of the parameter $\kappa$: $\kappa=1$ and $\kappa=2$
respectively. In both figures, one can see that there is a low
temperature range where the behavior at zero temperature is
reproduced. Increasing the temperature, there is a zone with a
more complicated dynamics, and in all cases one can see that for
very high temperature the survival probability approaches unity.
This is a universal feature of the dynamics related to this
dissipative \LZ model, as we will show in the following.

\subsection{Explanation of the behavior at high Temperature}

In the high-temperature limit we have $N(\omega)\approx \Theta/\omega$, so that the rate equations can be written down as follows:
\begin{widetext}
\begin{subequations}
\begin{align}
\label{eq:rho11} \dot{\rho}_{11} &= i\Omega\rho_{12}-i\Omega\rho_{21}\,, \\
\dot{\rho}_{12} &= \left(i\Omega-\frac{\Gamma \Theta\epsilon \sin 2\varphi}{\epsilon^2-\omega_3^2}\right)\rho_{11}+ \left[i\Delta +\frac{\Gamma \Theta (\epsilon\cos 2\varphi - \omega_3)}{\epsilon^2-\omega_3^2}\right]\rho_{12} -i\Omega \rho_{22} + \frac{\Gamma \Theta \epsilon \sin 2\varphi}{\epsilon^2-\omega_3^2}\rho_{33}\,, \label{eq:rho12}\\
\dot{\rho}_{13}&= \left[i(\omega_3+\frac{\Delta}{2}) -\Gamma + \frac{\Gamma \Theta(\epsilon \cos 2\varphi-\omega_3)}{2(\epsilon^2-\omega_3^2)}\right]\rho_{13}
-i\Omega \rho_{23} + \frac{\Gamma \Theta \epsilon \sin 2\varphi}{\epsilon^2-\omega_3^2}\rho_{32}\,, \label{eq:rho13}\\
\dot{\rho}_{22} &= \left(-i\Omega-\frac{\Gamma \Theta\epsilon \sin 2\varphi}{\epsilon^2-\omega_3^2}\right)\rho_{12} +\left(i\Omega-\frac{\Gamma \Theta\epsilon \sin 2\varphi}{\epsilon^2-\omega_3^2}\right)\rho_{21} +\frac{2\Gamma \Theta\left(-\omega_3+\epsilon \cos 2\varphi\right)}{\epsilon^2-\omega_3^2}\rho_{22}\nonumber\\
 &\quad + 2\Gamma\left(1+\Theta\frac{\omega_3-\epsilon \cos 2\varphi}{\epsilon^2-\omega_3^2}\right)\rho_{33}\,, \label{eq:rho22}\\
\dot{\rho}_{23}&= \left(-i\Omega-\frac{\Gamma \Theta\epsilon \sin 2\varphi}{\epsilon^2-\omega_3^2}\right)\rho_{13} +\left[i(\omega_3-\frac{\Delta}{2}) -\Gamma + \frac{\Gamma \Theta(\epsilon \cos 2\varphi-\omega_3)}{\epsilon^2-\omega_3^2}\right]\rho_{23} +\frac{\Gamma \Theta\epsilon \sin 2\varphi}{\epsilon^2-\omega_3^2}\rho_{31}\nonumber\\
&\quad +\Gamma\left(1+2\Theta\frac{\omega_3-\epsilon \cos 2\varphi}{\epsilon^2-\omega_3^2}\right)\rho_{32}\,, \label{eq:rho23} \\
 \dot{\rho}_{33}&= \frac{\Gamma \Theta\epsilon \sin 2\varphi}{\epsilon^2-\omega_3^2}\rho_{12} +\frac{\Gamma\Theta\epsilon \sin 2\varphi}{\epsilon^2-\omega_3^2}\rho_{21}
 -\frac{2\Gamma \Theta\left(-\omega_3+\epsilon \cos 2\varphi\right)}{\epsilon^2-\omega_3^2}\rho_{22}-2\Gamma\left(1+\Theta\frac{\omega_3-\epsilon \cos 2\varphi}{\epsilon^2-\omega_3^2}\right)\rho_{33}\,,\label{eq:rho33}
\end{align}
\end{subequations}
and $\dot{\rho}_{21}=\dot{\rho}_{12}^*$, $\dot{\rho}_{31}=\dot{\rho}_{13}^*$, $\dot{\rho}_{32}=\dot{\rho}_{23}^*$.
Let us introduce the perturbation parameters $\eta=\omega_3/\Theta \ll 1$ and $\xi = \epsilon/\omega_3 \ll 1$ (recall that $\Omega
\le \epsilon$ and $\Delta \le \epsilon$) and define the symbol $O_x\equiv O(x)$, i.e. infinitesimal of the order of $x$ or smaller.
On this basis, after introducing the vector notation for the density operator $\vec{\rho}=(\rho_{11}, \rho_{12}, \rho_{13}, \rho_{21}, \rho_{22}, \rho_{23}, \rho_{31}, \rho_{32}, \rho_{33})$, one can rewrite the rate equations (\ref{eq:rho11})-(\ref{eq:rho33}) as $\mathrm{d}\vec{\rho}/\mathrm{d}t={\cal L}\vec{\rho}$, with
\begin{equation}
{\cal L} = \Theta\, \left( \begin{array}{ccccccccc}
0 & O_{\xi\eta} & 0 & O_{\xi\eta} & 0 & 0 & 0 & 0 & 0 \\
O_{\xi\eta}+O_{\xi^2} & O_{\xi} & 0 & 0 & O_{\xi\eta} & 0 & 0 & 0 & O_{\xi^2} \\
0 & 0 & O_{\xi}+O_{\eta} & 0 & 0 & O_{\xi\eta} & 0 & O_{\xi^2} & 0 \\
O_{\xi\eta}+O_{\xi^2} & 0 & 0 & O_{\xi} & O_{\xi\eta} & 0 & 0 & 0 & 0 \\
0 & O_{\xi\eta}+O_{\xi^2} & 0 & O_{\xi\eta}+O_{\xi^2} & O_{\xi} & 0 & 0 & 0 & O_{\xi} \\
0 & 0 & O_{\xi\eta}+O_{\xi^2} & 0 & 0 & O_{\xi}+O_{\eta} & O_{\xi^2} & O_{\xi} & 0 \\
0 & 0 & 0 & 0 & 0 & O_{\xi^2} & O_{\xi}+O_{\eta} & O_{\xi\eta} & 0 \\
0 & 0 & O_{\xi^2} & 0 & 0 & O_{\xi} & O_{\xi\eta}+O_{\xi^2} & O_{\xi}+O_{\eta} & 0 \\
0 & O_{\xi^2} & 0 & O_{\xi^2} & O_{\xi} & 0 & 0 & 0 & O_{\xi} \\
\end{array} \right)\,,
\end{equation}
where we have used such equivalences as $\Omega/\Theta = \Omega/\omega_3 \times \omega_3/\Theta = O(\xi\eta)$, etc.
Keeping only first order terms in $\xi$ and $\eta$, one obtains:
\begin{equation}
{\cal L} = \Theta\, \left(
\begin{array}{ccccccccc}
0 & 0 & 0 & 0 & 0 & 0 & 0 & 0 & 0 \\
0 & O_{\xi} & 0 & 0 & 0 & 0 & 0 & 0 & 0 \\
0 & 0 & O_{\xi}+O_{\eta} & 0 & 0 & 0 & 0 & 0 & 0 \\
0 & 0 & 0 & O_{\xi} & 0 & 0 & 0 & 0 & 0 \\
0 & 0 & 0 & 0 & O_{\xi} & 0 & 0 & 0 & O_{\xi} \\
0 & 0 & 0 & 0 & 0 & O_{\xi}+O_{\eta} & 0 & O_{\xi} & 0 \\
0 & 0 & 0 & 0 & 0 & 0 & O_{\xi}+O_{\eta} & 0 & 0 \\
0 & 0 & 0 & 0 & 0 & O_{\xi} & 0 & O_{\xi}+O_{\eta} & 0 \\
0 & 0 & 0 & 0 & O_{\xi} & 0 & 0 & 0 & O_{\xi} \\
\end{array} \right)\,,
\end{equation}
\end{widetext}
from which we conclude that the projector $\Ket{1}\Bra{1}$ is an eigenoperator of the Lindbladian corresponding to the eigenvalue $0$, at first order in $\eta$ and $\xi$.
This explains why the population of the state $\Ket{1}$ is preserved at high temperature.
In fact, we are in the presence of a temperature-induced dynamical decoupling (or temperature-induced quantum Zeno effect).
Though this effect resembles the dynamical decoupling induced by decay \cite{ref:Militello2011-PhysicaScripta, ref:PascazioFacchi2001},
 it is a quite different phenomenon since it derives physically from the presence of thermal photons and the ensuing growth of decay and pumping rates at very high temperature.

It is important to note that the factor $\Theta$ does affect the eigenvalues of the Lindbladian, but it does not affect the structure of the eigenstates.

\section{Conclusions}

In this paper we have analyzed the Landau-Zener-St\"uckelberg-Majorana model in the presence of a dissipative bosonic enviroment
 by using a microscopic approach to derive the relevant master equation.
Because of the presence of a level crossing, the rotating-wave approximation cannot be made for the system-environment coupling.
The time evolution predicted by our microscopic model at zero temperature and weak dissipation is nearly identical to the time evolution predicted by the well-known phenomenological approach or the original \LZ model (assuming a constant coupling of infinite duration) \cite{ref:Vitanov97}: the survival probability of state $\Ket{1}$ is independent of the decay rate.
For strong dissipation we find that the initial-state population approaches unity, as in the finite \LZ model \cite{ref:Vitanov96} and other analytic models with couplings of finite duration \cite{ref:Vitanov97}.

An important advantage of our microscopic approach is that it can describe also the dissipative dynamics for an environment at a nonzero temperature.
A distinct feature in this case is the survival probability approach to unity for very high temperatures.
The physical origin of this effect is the dynamical decoupling between state $\Ket{1}$ and the subspace spanned by states $\Ket{2}$ and $\Ket{3}$.
This is induced by the very strong coupling between states $\Ket{2}$ and $\Ket{3}$ mediated by the interaction with the environment, which becomes stronger as the number of photons increases.
It is important to note that in order to verify these results the initial state ($\Ket{1}$) must not be electromagnetically coupled to the external state ($\Ket{3}$),
 so that all dissipation is due to state $\Ket{2}$, as in Ref.~\cite{ref:Vitanov97}.
Therefore this very peculiar feature of the high-temperature dynamics should be measurable in experimental situations where only one of the two states is unstable.

\section*{Acknowledgements}
This work is supported by the MIUR Project N. II04C0E3F3, the European Commission's project FASTQUAST, and
the Bulgarian NSF grant D002-90/08.


\begin{thebibliography}{99}

\bibitem{ref:LZSM-1} L. D. Landau, Physik Z. Sowjetunion \textbf{2}, 46 (1932).

\bibitem{ref:LZSM-2} C. Zener, Proc. R. Soc. Lond. Ser. A \textbf{137}, 696 (1932).

\bibitem{ref:LZSM-3} E. C. G. St\"{u}ckelberg, Helv. Phys. Acta \textbf{5}, 369 (1932).

\bibitem{ref:LZSM-4} E. Majorana, Nuovo Cimento \textbf{9}, 43 (1932).

\bibitem {ref:ARPC} N.V. Vitanov, T. Halfmann, B. W. Shore, and K. Bergmann, Ann. Rev. Phys. Chem. \textbf{52}, 763 (2001).

\bibitem{ref:Ao89} P. Ao and J. Rammer, Phys. Rev. Lett. \textbf{62}, 3004 (1989); Phys. Rev. B \textbf{43}, 5397 (1991).

\bibitem{ref:Akulin} V. M. Akulin and W. P. Schleich, Phys. Rev. A \textbf{46}, 4110 (1992).

\bibitem{ref:Vitanov97} N. V. Vitanov and S. Stenholm, Phys. Rev. A \textbf{55}, 2982 (1997).

\bibitem{ref:Vitanov96} N. V. Vitanov and B. M. Garraway, Phys. Rev. A \textbf{53}, 4288 (1996);
erratum ibid. \textbf{54}, 5458 (1997).

\bibitem{ref:Vitanov99} N. V. Vitanov, Phys. Rev. A \textbf{59}, 988 (1999).

\bibitem {ref:Nelbach} P. Nalbach and M. Thorwart, Phys. Rev. Lett \textbf{103}, 220401 (2009).

\bibitem {ref:Orth} P. P. Orth, A. Imambekov, and K. Le Hur, Phys. Rev. A \textbf{82}, 032118 (2010).

\bibitem{ref:DDP} J. P. Davis and P. Pechukas, J. Chem. Phys. \textbf{64}, 3129 (1976).

\bibitem{ref:Davies1978} E. B. Davies and H. Spohn, J. Stat. Phys. \textbf{19}, 511 (1978).

\bibitem{ref:Scala2010} M. Scala, B. Militello, A. Messina, and N. V. Vitanov, Phys. Rev. A {\bf 81}, 053847 (2010).

\bibitem{ref:Scala2011} M. Scala, B. Militello, A. Messina, and N. V. Vitanov, Phys. Rev. A {\bf 83}, 012101 (2011).

\bibitem{ref:Nori_review} S.-N. Shevchenko, S. Ashahab, and F. Nori, Phys. Rep. {\bf 492}, 1 (2010).

\bibitem{ref:JJ1} D.M. Berns, {\it et al.}, Nature (London) {\bf 455}, 51 (2008).

\bibitem{ref:JJ2} G. Sun, {\it et al.}, Nature Commun. {\bf 1}, 51 (2010).

\bibitem{ref:BEC} A. Zenesini, D. Ciampini, O. Morsch, and E. Arimondo, Phys. Rev. A {\bf 82}, 065601 (2010).

\bibitem{ref:Nori_spinorial} J.-N. Zhang, C.-P. Sun, S. Yi, and F. Nori, Phys. Rev. A {\bf 83}, 033614 (2011).

\bibitem{ref:Florio2006} G. Florio, P. Facchi, R. Fazio, V. Giovannetti, and S.Pascazio, Phys. Rev. A {\bf 73}, 022327 (2006).

\bibitem{ref:Gardiner} C.~W. Gardiner and P. Zoller, {\it Quantum Noise\/} (Springer-Verlag, Berlin, 2000).

\bibitem{ref:Petru} H.-P. Breuer and F. Petruccione, {\it The Theory of Open Quantum Systems\/} (Oxford University Press, Oxford, 2002).

\bibitem{ref:Militello2011-PhysicaScripta} B. Militello, M. Scala, A. Messina, and N. V. Vitanov, Phys. Scr. {\bf T143}, 014019 (2011).

\bibitem{ref:PascazioFacchi2001} P. Facchi and S. Pascazio, {\sl Progress in Optics} {\bf 41} edited by E. Wolf, Elsevier, Amsterdam, 2001.

\end{thebibliography}
\end{document}